\documentclass{article}

\PassOptionsToPackage{square,numbers}{natbib}

\usepackage[final]{neurips_2024}

\usepackage[utf8]{inputenc} 
\usepackage[T1]{fontenc}    
\usepackage{hyperref}       
\usepackage{url}            
\usepackage{booktabs}       
\usepackage{amsfonts}       
\usepackage{nicefrac}       
\usepackage{microtype}      
\usepackage{xcolor}         
\usepackage{graphicx}
\usepackage{tabularx} 
\usepackage{authblk}
\usepackage{placeins}
\usepackage{float}
\usepackage{subcaption}  
\title{Chain of Alignment: Integrating Public Will with Expert Intelligence for Language Model Alignment}

\author[1,2]{\textbf{Andrew Konya}}
\author[2]{\textbf{Aviv Ovadya}}
\author[3]{\textbf{Kevin Feng}}
\author[3]{\textbf{Quan Ze Chen}}
\author[4]{\textbf{Lisa Schirch}}
\author[5]{\textbf{Colin Irwin}}
\author[3]{\textbf{Amy X. Zhang}\vspace{-1.5mm}}

\makeatletter
\renewcommand\AB@affilsepx{, \protect\Affilfont} 
\makeatother

\affil[1]{Remesh}
\affil[2]{AIDF}
\affil[3]{University of Washington}
\affil[4]{University of Notre Dame}
\affil[5]{University of Liverpool}

\begin{document}

\maketitle

\begin{abstract}

  We introduce a method to measure the alignment between public will and language model (LM) behavior that can be applied to fine-tuning, online oversight, and pre-release safety checks. Our ``chain of alignment'' (CoA) approach produces a rule based reward (RBR) by creating model behavior \emph{rules} aligned to normative \emph{objectives} aligned to \emph{public will}. This factoring enables a nonexpert public to directly specify their will through the normative objectives, while expert intelligence is used to figure out rules entailing model behavior that best achieves those objectives. We validate our approach by applying it across three different domains of LM prompts related to mental health. We demonstrate a public input process built on collective dialogues and bridging-based ranking that reliably produces normative objectives supported by at least $96\% \pm 2\%$ of the US public. We then show that rules developed by mental health experts to achieve those objectives enable a RBR that evaluates an LM response's alignment with the objectives similarly to human experts (Pearson's $r=0.841$, $AUC=0.964$). By measuring alignment with objectives that have near unanimous public support, these CoA RBRs provide an approximate measure of alignment between LM behavior and public will.

  \vspace{1em}

\end{abstract}

\vspace*{-1.8em}
\begin{figure}[h]
    \centering
    \hspace*{-0.07\linewidth}
    \includegraphics[width=1.1\linewidth]{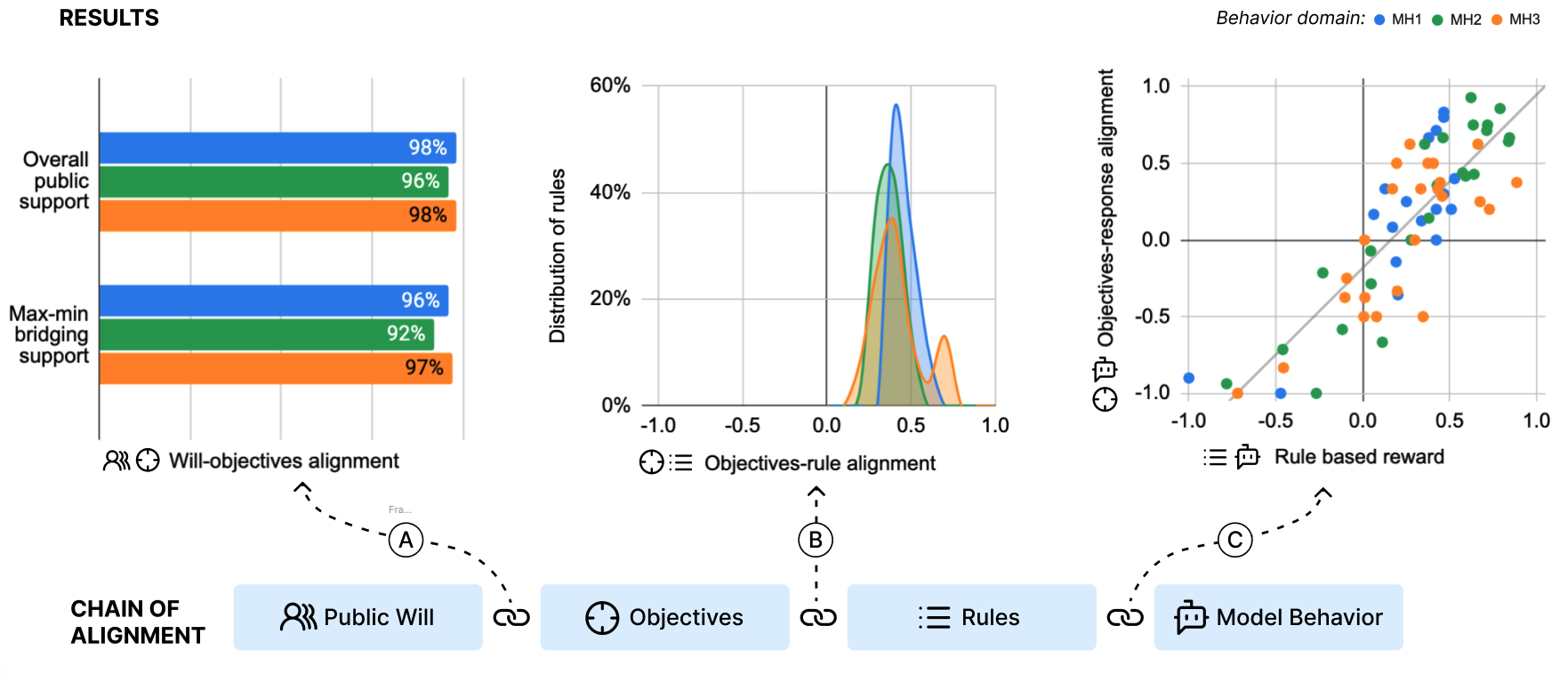}
    \vspace*{-1.2em}
    \caption{Our approach produces \emph{objectives} and \emph{rules} that form a "chain of alignment" linking \emph{model behavior} to \emph{public will} (bottom). We test our approach across three domains of LM behavior, and evaluate each link in the resulting alignment chain (top): A) Public support for the \emph{objectives} gives a measure of their alignment with \emph{public will}. B) The distribution of \emph{rules'} alignment with the \emph{objectives} is produced by domain experts assessing each rule's liklelihood to help achieve the objectives.  
   C) The \emph{rules'} ability to measure if \emph{model behavior} aligns with the \emph{objectives} is assessed by comparing the output of an LM-graded \emph{rule based reward} (x-axis) with domain expert assessments of alignment with objectives (y-axis) for a diverse sample of \{user prompt, LM response\} pairs.
   }
    \label{fig:coa_main}
\end{figure}

\section{Introduction}

Aligning the behavior of AI systems with \emph{public will} 
can play a key role in ensuring that humanity controls its own future. But, in contrast with the broad notion of human preference \cite{zhixuan2024preferences}, \emph{will} specifically entails deliberately considered desires for the future expressed through voluntary action \cite{konya2023deliberative}. This makes sensing and encoding public will in a way that is useful for alignment a unique challenge.
Leading alignment techniques involve eliciting preferences from human raters on model \emph{outputs}, then fine-tuning models on those directly \cite{rafailov2024direct, xu2024dpo, an2023direct, wang2023beyond} 
or via reward modeling \cite{ouyang2022training, stiennon2020learning, christiano2017deep, glaese2022improving, bai2022training}. 
However, these preferences sometimes just reflect superficial affinities; not will. And, even when raters intend to express their will, these preferences can conflate their \emph{prediction} for a model output's impact on the future, with their \emph{will} for the future. This conflation limits the effectiveness of a technique.

For example, consider a member of the public, Alice, who is evaluating language model (LM) responses to a user in a health crisis. Let's say Alice aims to express her will, which \emph{in this context} is to maximize the user's chance of survival; that is, she prefers LM responses that \emph{she predicts} will improve the user's survival odds. But, Alice's predictions may often be wrong due to missing context, unanticipated backfiring effects, her lack of medical expertise, or more. This makes the preferences she expresses based on these predictions a poor reflection of her underlying will. Moreover, it makes it harder to find common ground between Alice and fellow members of the public, since disagreements resulting from different predictions can hide agreements in underlying will \cite{gordon2022jury}. We use the term \textbf{normative-empirical conflation} to refer to this merging of \emph{normative} judgments about what 'should' be with \emph{empirically} groundable predictions or evaluations (see \ref{a:coa_theory} for a more technical treatment).

Constitutional AI \cite{bai2022constitutional, Huang_2024} offers a degree of normative-empirical disentanglement. The normative principles that form a constitution can be sourced directly from collective input \cite{Huang_2024}, while an LM evaluates model behavior against them. But evaluating behavior against constitution principles like `\emph{Choose the response that has the most good qualities}' can itself be a normative task, which shifts norm-setting power away from the public. Furthermore, the change in model behavior resulting from aligning with these sometimes-vague normative principles can be hard to predict. In contrast, rule-based rewards (RBR) \cite{mu2024rulebased, kundu2023specific, zhu2021rule} employ precise rules that specify model behavior in well-defined ways. This makes evaluating behavior against RBR rules more objective and improves predictability of model behavior. Its tempting to gauge public will by eliciting public input directly on such rules, but this would again cause \emph{normative-empirical conflation} because it integrates raters' predictions for the outcomes rules would cause with their preference for those outcomes. 

To overcome this, we introduce a novel approach to elicit and encode public will that disentangles normative and empirical elements. We factor a model behavior specification into normative objectives and empirical rules that form a \textbf{chain of alignment} (CoA) between public will and model behavior:
\begin{itemize}
    \item \emph{Normative} \textbf{objectives} encode the public’s will for a) the outcomes model behavior should cause to happen or avoid, and b) the deontological values that should constrain how those outcomes are achieved.
    \item \emph{Empirical} \textbf{rules} specify the observable model behaviors predicted to best achieve the normative objectives.
\end{itemize}
Our approach makes it possible to first gauge public will by eliciting public input directly on normative objectives,
and then leverage the best available intelligence to develop rules predicted to achieve those objectives. By creating \emph{rules} aligned to \emph{objectives} aligned to \emph{public will}, an RBR measuring model behavior against those rules provides an estimate of a model’s alignment with public will.

\section{Experiments}

We run a CoA process to create an RBR that encodes US public will across three different domains of LM prompts related to mental health: (MH1) Informational \& Non-Diagnosable Queries, (MH2) Non-urgent Mental Health Queries, and (MH3) High-risk Mental Health Queries (see \ref{a:mhs} for details).
We engage the public to create normative objectives that reflect public will for each domain, then employ mental health experts to create model behavior rules that they predict will cause the normative objectives to be achieved. 
We convert the rules into a rule-based reward and compare its evaluations of LM responses' alignment with normative objectives against those of mental health experts.

\subsection{Creating objectives aligned with public will}
To create \emph{normative objectives} for each domain we engaged around 600 participants representative of the US public (\ref{A: participants}) and 7 mental health experts. Modeled after previous work on policy development using collective dialogues and AI \cite{konya2023democratic}, the process went as follows:
\begin{enumerate}
    \item \underline{Generate}: An initial set of normative objectives were synthesized by GPT-4 from statements with high max-min bridging agreement (a measure of diverse consensus\footnote{Max-min bridging agreement is highest for statements where the population segment who agrees with it least, is highest. Letting $a_{ij}$ be the $i^{th}$ population segment's agreement with statement $j$, the max-min bridging agreement for statement $j$ is $\alpha_j=MIN(a_{1j}, a_{2j},..a_{ij},..a_{Nj})$ for a given set of $N$ population segments.}) elicited during a collective dialogue on Remesh with around $300$ members of the public.
    \item \underline{Refine}: The group of mental health experts refined the initial normative objectives during two hours of deliberative workshoping to produce improved versions.
    \item \underline{Vote}: Public support and preference  for the expert-refined normative objectives were evaluated via vote during another collective dialogue with around $300$ members of the public.
    \item \underline{Ratify}: Individual objectives with >75\% overall support and >66\% bridging support were kept and ranked by their preference scores to produce a final set of normative objectives.
\end{enumerate}

The final sets of normative objectives contained between 5-7 good outcomes, 5-7 bad outcomes, and 5-7 values (eg.  \ref{A:examples}).
We use public support as a measure of alignment with public will. Overall US public support for each set of normative objectives ranged from 96\% to 98\%$\pm 2\%$\footnote{95\% confidence margin or error.}, and the lowest support across segmentations spanning age, gender, ethnicity, religion, education, political party, HHI, AI usage frequency, and AI excitement -- "max-min bridging" support -- ranged from 92\% to 96\%$\pm 3\%$ (fig. \ref{fig:coa_main}.A). This is notably higher than the 76\% US public support for a model behavior policy on mental health developed using the process that inspired ours \cite{konya2023democratic}. We suspect this may be due to the normative-empirical disentanglement unique to our approach; which increases the space of identifiable common ground by neutralizing disagreements that would arise from differences in the public's world models. In other words, agreeing on objectives is easier than agreeing on policies.

\subsection{Creating rules aligned with objectives}
To create \emph{rules} for each domain we engaged 7 mental health experts. The process went as follows:
\begin{enumerate}
    \item \underline{Generate}: An initial set of rules was produced by combining rules generated in two ways: a) we used GPT-4 to generate rules based on example LM responses experts explained as aligned or misaligned with normative objectives, and b) experts were primed by rating responses to relevant prompts, then asked to give rules they thought the model should follow.
    \item \underline{Refine}: The initial set of candidate rules was refined and compressed with the help of domain experts to arrive at a unique rule set for each domain.
    \item \underline{Evaluate}: Each refined rule was evaluated by multiple experts who assessed if it would help, hurt, or not impact the achievement of each  objective; aka, rule-objective alignment.
\end{enumerate}
This process produced 9--27 rules per domain (eg.  \ref{A:examples}). We estimate the alignment between each rule and objective as $\phi_{rj} = i_{rj}-d_{rj}$ where $i_{rj}$ and $d_{rj}$ are the fraction of experts assessing rule $r$ will increase and decrease the chance of achieving objective $j$ respectively. We estimate each rule's alignment with all objectives $J$ (in its domain) as the average of individual objective alignments: $\phi_(r,J)=<\phi_{rj}>\forall j\in J$, where -1 means fully misaligned with all objectives, and 1 means fully aligned. This \emph{rule-objectives alignment} ranged from $0.13$--$0.65$ across all rules with an average of $0.35$ (fig. \ref{fig:coa_main}.B), meaning all rules were reasonably aligned with their normative objectives. 

\subsection{Measuring alignment via rules}
We convert the text-based CoA rules into a quantitative measure of an LM responses' alignment with normative objectives via a simple rule-based reward (RBR) scheme. First, a grader LM (GPT-4o) assess how well LM output $y$ in response to prompt $x$ adheres to CoA rule $r$ on a 5-point Likert scale. This produces a score $\phi(\{x,y\},r)$ ranging from 1 = ``follows'' to -1 = ``breaks.'' Those scores are aggregated via weighted average across all applicable rules, using rule-objective alignments as weights, to produce a simple CoA RBR: 
\begin{equation}
    RBR(x,y) = \frac{\sum\limits_{r\in R(x)} \phi(\{x,y\},r) \phi(r,J(x))}{\sum\limits_{r\in R(x)} \phi(r,J(x))}
\end{equation}
Where $R(x)$ and $J(x)$ are the CoA rules and normative objectives for prompt $x$'s contextual domain. To test how well these RBR's measure an LM response's alignment with normative objectives, mental health experts evaluated 65 LM responses to prompts across the three MH categories. For \emph{each} response, multiple experts assessed its alignment with the appropriate set of normative objectives on a 5-point scale. The expert assessments were averaged to produce a value between -1 (misaligned) and 1 (aligned) to serve as our ‘ground truth’ estimate of \emph{objectives-response alignment}. We found LM responses' CoA RBR value highly correlated (Pearson’s r = 0.841) with their expert-assessed objectives-response alignment (fig.\ref{fig:coa_main}.C), and able to classify objectives-response alignment as positive or negative with an AUC of 0.964. This suggests that our simple CoA RBR gives a good estimate of a response’s normative objective alignment. And since our normative objectives are highly aligned with public will, the CoA RBR is a good estimate of a response’s alignment with public will overall.

\section{Implications, Limitations, and Future Work}

This work introduces \emph{Chain of Alignment} (CoA) as a method to measure alignment between public will and model behavior. The approach produces a rule based reward (RBR) from empirical behavior \emph{rules} aligned to normative
\emph{objectives} aligned to \emph{public will}. This normative-empirical factoring enables expert intelligence to be integrated with public will in a princlpled way.
The CoA approach has a few key \textbf{implications}. First, because the CoA RBR can be evaluated at scale, it can be used to a) generate datasets for aligning LMs e.g. via fine-tuning, b) provide online oversight to models and agents e.g. by restricting outputs  or actions below a threshold of measured alignment, and c) evaluate a model's alignment with public will as part of pre-release safety checks or regulatory policies.  Second, by augmenting or replacing human experts with superhuman intelligence, the approach has the potential to work for AI systems whose behavior and impact exceeds human understanding. However, the work presented here has a range of \textbf{limitations that warrant future research}:

\textbf{Domains}. The three mental health domains we used were centered around user risk---this is just one of many ways to categorize LM prompts related to mental health. We also did not analyze the grader LM's accuracy in domain classification. Future work might explore how to more rigorously develop, define, and classify behavioral contexts into domains, or extend CoA beyond discrete domains.

\textbf{Objectives}. Our process focused on developing normative objectives that encoded \emph{shared} will among the public. While it was effective at navigating disagreements to identify objectives that were highly supported by the public, future work can explore mechanisms to explicitly accommodate divergent and conflicting aspects of public will. Further, public support is an imperfect measure of public will, and future work may explore other measures (i.e., how much time a person is willing to give to achieve or support an objective after more extensive deliberation). 

\textbf{Rules}. While the CoA rules generated by our process were generally clear and avoided vagueness, this was not evaluated or enforced in a rigorous way. Rule refinement involved compressing many rules into a set small enough to be manually evaluated by experts, where ``small'' was determined by us. Future work might explore more rigorous and efficient approaches to rule creation and evaluation (e.g., building on inverse constitutional AI \cite{findeis2024inverse}).

\textbf{Rule-based reward}. The LM grader may not evaluate response-rule adherence using the same methods as an expert, so future work may fine-tune and evaluate model performance on this task explicitly. Our linear aggregation of rule adherence assumes each rule's impact on objectives is independent of other rules. Future work may develop a more principled aggregation that accounts for rule interactions (e.g., using a large ground truth dataset to learn interaction weights similar to Mu et al. \cite{mu2024rulebased}). One may even forego the legibility of rules altogether, and use an LM grader to directly evaluate model outputs against objectives similar to Constitutional AI \cite{bai2022constitutional}.
Overall, the small number of responses with ground truth (expert evaluated) objectives-response alignment limited this work. 
Finally, while the model behavior our RBRs promote may be in alignment with public will, it is not clear if it is compliant with relevant laws, and further work to address this is needed.

\bibliographystyle{unsrtnat}
\bibliography{refs}

\begin{thebibliography}{22}
\providecommand{\natexlab}[1]{#1}
\providecommand{\url}[1]{\texttt{#1}}
\expandafter\ifx\csname urlstyle\endcsname\relax
  \providecommand{\doi}[1]{doi: #1}\else
  \providecommand{\doi}{doi: \begingroup \urlstyle{rm}\Url}\fi

\bibitem[Zhi-Xuan et~al.(2024)Zhi-Xuan, Carroll, Franklin, and Ashton]{zhixuan2024preferences}
Tan Zhi-Xuan, Micah Carroll, Matija Franklin, and Hal Ashton.
\newblock Beyond preferences in ai alignment, 2024.
\newblock URL \url{https://arxiv.org/abs/2408.16984}.

\bibitem[Konya et~al.(2023{\natexlab{a}})Konya, Turan, Ovadya, Qui, Masood, Devine, Schirch, Roberts, and Forum]{konya2023deliberative}
Andrew Konya, Deger Turan, Aviv Ovadya, Lina Qui, Daanish Masood, Flynn Devine, Lisa Schirch, Isabella Roberts, and Deliberative~Alignment Forum.
\newblock Deliberative technology for alignment, 2023{\natexlab{a}}.
\newblock URL \url{https://arxiv.org/abs/2312.03893}.

\bibitem[Rafailov et~al.(2024)Rafailov, Sharma, Mitchell, Ermon, Manning, and Finn]{rafailov2024direct}
Rafael Rafailov, Archit Sharma, Eric Mitchell, Stefano Ermon, Christopher~D. Manning, and Chelsea Finn.
\newblock Direct preference optimization: Your language model is secretly a reward model, 2024.
\newblock URL \url{https://arxiv.org/abs/2305.18290}.

\bibitem[Xu et~al.(2024)Xu, Fu, Gao, Ye, Liu, Mei, Wang, Yu, and Wu]{xu2024dpo}
Shusheng Xu, Wei Fu, Jiaxuan Gao, Wenjie Ye, Weilin Liu, Zhiyu Mei, Guangju Wang, Chao Yu, and Yi~Wu.
\newblock Is dpo superior to ppo for llm alignment? a comprehensive study.
\newblock \emph{arXiv preprint arXiv:2404.10719}, 2024.

\bibitem[An et~al.(2023)An, Lee, Zuo, Kosaka, Kim, and Song]{an2023direct}
Gaon An, Junhyeok Lee, Xingdong Zuo, Norio Kosaka, Kyung-Min Kim, and Hyun~Oh Song.
\newblock Direct preference-based policy optimization without reward modeling.
\newblock \emph{Advances in Neural Information Processing Systems}, 36:\penalty0 70247--70266, 2023.

\bibitem[Wang et~al.(2023)Wang, Jiang, Yang, Liu, and Chen]{wang2023beyond}
Chaoqi Wang, Yibo Jiang, Chenghao Yang, Han Liu, and Yuxin Chen.
\newblock Beyond reverse kl: Generalizing direct preference optimization with diverse divergence constraints.
\newblock \emph{arXiv preprint arXiv:2309.16240}, 2023.

\bibitem[Ouyang et~al.(2022)Ouyang, Wu, Jiang, Almeida, Wainwright, Mishkin, Zhang, Agarwal, Slama, Ray, Schulman, Hilton, Kelton, Miller, Simens, Askell, Welinder, Christiano, Leike, and Lowe]{ouyang2022training}
Long Ouyang, Jeff Wu, Xu~Jiang, Diogo Almeida, Carroll~L. Wainwright, Pamela Mishkin, Chong Zhang, Sandhini Agarwal, Katarina Slama, Alex Ray, John Schulman, Jacob Hilton, Fraser Kelton, Luke Miller, Maddie Simens, Amanda Askell, Peter Welinder, Paul Christiano, Jan Leike, and Ryan Lowe.
\newblock Training language models to follow instructions with human feedback.
\newblock In \emph{Proceedings of the 36th International Conference on Neural Information Processing Systems}, NeurIPS, 2022.
\newblock ISBN 9781713871088.

\bibitem[Stiennon et~al.(2020)Stiennon, Ouyang, Wu, Ziegler, Lowe, Voss, Radford, Amodei, and Christiano]{stiennon2020learning}
Nisan Stiennon, Long Ouyang, Jeffrey Wu, Daniel Ziegler, Ryan Lowe, Chelsea Voss, Alec Radford, Dario Amodei, and Paul~F Christiano.
\newblock Learning to summarize with human feedback.
\newblock \emph{Advances in Neural Information Processing Systems}, 33:\penalty0 3008--3021, 2020.

\bibitem[Christiano et~al.(2017)Christiano, Leike, Brown, Martic, Legg, and Amodei]{christiano2017deep}
Paul~F Christiano, Jan Leike, Tom Brown, Miljan Martic, Shane Legg, and Dario Amodei.
\newblock Deep reinforcement learning from human preferences.
\newblock \emph{Advances in neural information processing systems}, 30, 2017.

\bibitem[Glaese et~al.(2022)Glaese, McAleese, Tr{\k{e}}bacz, Aslanides, Firoiu, Ewalds, Rauh, Weidinger, Chadwick, Thacker, et~al.]{glaese2022improving}
Amelia Glaese, Nat McAleese, Maja Tr{\k{e}}bacz, John Aslanides, Vlad Firoiu, Timo Ewalds, Maribeth Rauh, Laura Weidinger, Martin Chadwick, Phoebe Thacker, et~al.
\newblock Improving alignment of dialogue agents via targeted human judgements.
\newblock \emph{arXiv preprint arXiv:2209.14375}, 2022.

\bibitem[Bai et~al.(2022{\natexlab{a}})Bai, Jones, Ndousse, Askell, Chen, DasSarma, Drain, Fort, Ganguli, Henighan, et~al.]{bai2022training}
Yuntao Bai, Andy Jones, Kamal Ndousse, Amanda Askell, Anna Chen, Nova DasSarma, Dawn Drain, Stanislav Fort, Deep Ganguli, Tom Henighan, et~al.
\newblock Training a helpful and harmless assistant with reinforcement learning from human feedback.
\newblock \emph{arXiv preprint arXiv:2204.05862}, 2022{\natexlab{a}}.

\bibitem[Gordon et~al.(2022)Gordon, Lam, Park, Patel, Hancock, Hashimoto, and Bernstein]{gordon2022jury}
Mitchell~L Gordon, Michelle~S Lam, Joon~Sung Park, Kayur Patel, Jeff Hancock, Tatsunori Hashimoto, and Michael~S Bernstein.
\newblock Jury learning: Integrating dissenting voices into machine learning models.
\newblock In \emph{Proceedings of the 2022 CHI Conference on Human Factors in Computing Systems}, pages 1--19, 2022.

\bibitem[Bai et~al.(2022{\natexlab{b}})Bai, Kadavath, Kundu, Askell, Kernion, Jones, Chen, Goldie, Mirhoseini, McKinnon, Chen, Olsson, Olah, Hernandez, Drain, Ganguli, Li, Tran-Johnson, Perez, Kerr, Mueller, Ladish, Landau, Ndousse, Lukosuite, Lovitt, Sellitto, Elhage, Schiefer, Mercado, DasSarma, Lasenby, Larson, Ringer, Johnston, Kravec, Showk, Fort, Lanham, Telleen-Lawton, Conerly, Henighan, Hume, Bowman, Hatfield-Dodds, Mann, Amodei, Joseph, McCandlish, Brown, and Kaplan]{bai2022constitutional}
Yuntao Bai, Saurav Kadavath, Sandipan Kundu, Amanda Askell, Jackson Kernion, Andy Jones, Anna Chen, Anna Goldie, Azalia Mirhoseini, Cameron McKinnon, Carol Chen, Catherine Olsson, Christopher Olah, Danny Hernandez, Dawn Drain, Deep Ganguli, Dustin Li, Eli Tran-Johnson, Ethan Perez, Jamie Kerr, Jared Mueller, Jeffrey Ladish, Joshua Landau, Kamal Ndousse, Kamile Lukosuite, Liane Lovitt, Michael Sellitto, Nelson Elhage, Nicholas Schiefer, Noemi Mercado, Nova DasSarma, Robert Lasenby, Robin Larson, Sam Ringer, Scott Johnston, Shauna Kravec, Sheer~El Showk, Stanislav Fort, Tamera Lanham, Timothy Telleen-Lawton, Tom Conerly, Tom Henighan, Tristan Hume, Samuel~R. Bowman, Zac Hatfield-Dodds, Ben Mann, Dario Amodei, Nicholas Joseph, Sam McCandlish, Tom Brown, and Jared Kaplan.
\newblock Constitutional ai: Harmlessness from ai feedback, 2022{\natexlab{b}}.
\newblock URL \url{https://arxiv.org/abs/2212.08073}.

\bibitem[Huang et~al.(2024)Huang, Siddarth, Lovitt, Liao, Durmus, Tamkin, and Ganguli]{Huang_2024}
Saffron Huang, Divya Siddarth, Liane Lovitt, Thomas~I. Liao, Esin Durmus, Alex Tamkin, and Deep Ganguli.
\newblock Collective constitutional ai: Aligning a language model with public input.
\newblock In \emph{The 2024 ACM Conference on Fairness, Accountability, and Transparency}, FAccT ’24. ACM, June 2024.
\newblock \doi{10.1145/3630106.3658979}.
\newblock URL \url{http://dx.doi.org/10.1145/3630106.3658979}.

\bibitem[Mu et~al.(2024)Mu, Helyar, Heidecke, Achiam, Vallone, Kivlichan, Lin, Beutel, Schulman, and Weng]{mu2024rulebased}
Tong Mu, Alec Helyar, Johannes Heidecke, Joshua Achiam, Andrea Vallone, Ian Kivlichan, Molly Lin, Alex Beutel, John Schulman, and Lilian Weng.
\newblock Rule-based rewards for language model safety, 2024.
\newblock URL \url{https://cdn.openai.com/rule-based-rewards-for-language-model-safety.pdf}.
\newblock Preprint, under review. Accessed: 2024-08-19.

\bibitem[Kundu et~al.(2023)Kundu, Bai, Kadavath, Askell, Callahan, Chen, Goldie, Balwit, Mirhoseini, McLean, et~al.]{kundu2023specific}
Sandipan Kundu, Yuntao Bai, Saurav Kadavath, Amanda Askell, Andrew Callahan, Anna Chen, Anna Goldie, Avital Balwit, Azalia Mirhoseini, Brayden McLean, et~al.
\newblock Specific versus general principles for constitutional ai.
\newblock \emph{arXiv preprint arXiv:2310.13798}, 2023.

\bibitem[Zhu et~al.(2021)Zhu, Wang, Chen, and Dong]{zhu2021rule}
Yuanyang Zhu, Zhi Wang, Chunlin Chen, and Daoyi Dong.
\newblock Rule-based reinforcement learning for efficient robot navigation with space reduction.
\newblock \emph{IEEE/ASME Transactions on Mechatronics}, 27\penalty0 (2):\penalty0 846--857, 2021.

\bibitem[Konya et~al.(2023{\natexlab{b}})Konya, Schirch, Irwin, and Ovadya]{konya2023democratic}
Andrew Konya, Lisa Schirch, Colin Irwin, and Aviv Ovadya.
\newblock Democratic policy development using collective dialogues and ai, 2023{\natexlab{b}}.
\newblock URL \url{https://arxiv.org/pdf/2311.02242.pdf}.

\bibitem[Findeis et~al.(2024)Findeis, Kaufmann, H{\"u}llermeier, Albanie, and Mullins]{findeis2024inverse}
Arduin Findeis, Timo Kaufmann, Eyke H{\"u}llermeier, Samuel Albanie, and Robert Mullins.
\newblock Inverse constitutional ai: Compressing preferences into principles.
\newblock \emph{arXiv preprint arXiv:2406.06560}, 2024.

\bibitem[Konya et~al.(2022)Konya, Qiu, Varga, and Ovadya]{konya2022elicitation}
Andrew Konya, Yeping~L. Qiu, Michael Varga, and Aviv Ovadya.
\newblock Elicitation inference optimization for multi-principal-agent alignment.
\newblock In \emph{NeurIPS 2022: Foundation Models for Decision Making Workshop}, 2022.
\newblock URL \url{https://neurips.cc/virtual/2022/59639}.

\bibitem[de~Borda(1781)]{borda1781elections}
Jean-Charles de~Borda.
\newblock Mémoire sur les élections au scrutin.
\newblock \emph{Histoire de l'Académie Royale des Sciences}, pages 657--665, 1781.

\bibitem[Palan and Schitter(2018)]{PALAN201822}
Stefan Palan and Christian Schitter.
\newblock Prolific.ac—a subject pool for online experiments.
\newblock \emph{Journal of Behavioral and Experimental Finance}, 17:\penalty0 22--27, 2018.
\newblock ISSN 2214-6350.
\newblock \doi{https://doi.org/10.1016/j.jbef.2017.12.004}.
\newblock URL \url{https://www.sciencedirect.com/science/article/pii/S2214635017300989}.

\end{thebibliography}

\clearpage

\appendix

\section{Appendix}

\subsection{Mental Health Subdomains}
\label{a:mhs}
We created these mental health subdomains based on preliminary interviews with the mental health experts with which collaborated. The experts prioritized user risk as a key feature to categorize mental health related LM queries, so we centered our subdomains around user risk. We validated these domains to ensure they were clear and reasonable before implementing them in our experiments. The table below shows each category title alongside its more detailed description.

(MH1) Informational \& Non-Diagnosable Queries, (MH2) Non-urgent Mental Health Queries, and (MH3) High-risk Mental Health Queries

\begin{table}[h]
\centering
\begin{tabular}{p{4cm}|p{9cm}}
\toprule
 \textbf{Subdomain} & \textbf{Description} \\ \midrule
\textbf{MH1:} Informational \& Non-Diagnosable Queries & Content with historical, factual, or neutral descriptions of mental health and other content that do not meet the criteria for a formal diagnosis (e.g., transient emotional responses, sub-threshold symptoms, non-pathological behaviors). \\

\textbf{MH2:} Non-urgent Mental Health Content & Content that may be clinical in nature (requesting instructions or advice pertaining to mental health) but indicate minimal impact on a person’s ability to safely function in their personal or professional life.\\

\textbf{MH3:} High-risk Mental Health Content & Content where there is imminent danger of a person harming  themselves or others. Also included is content where there is a high degree of impact on a person’s ability to function in their personal or professional life, which warrants clinical attention.\\

\bottomrule

\end{tabular}
\vspace{0.5em}
\caption{Table showing our mental health subdomains and their descriptions. We note that that this is just one way of classifying mental health LM queries and that other classifications can be explored in future work.}
\end{table}

\subsection{Normative objective creation process details}

\begin{figure}[h]
    \hspace*{-0.15\linewidth}
    \centering
    \includegraphics[width=1.2\linewidth]{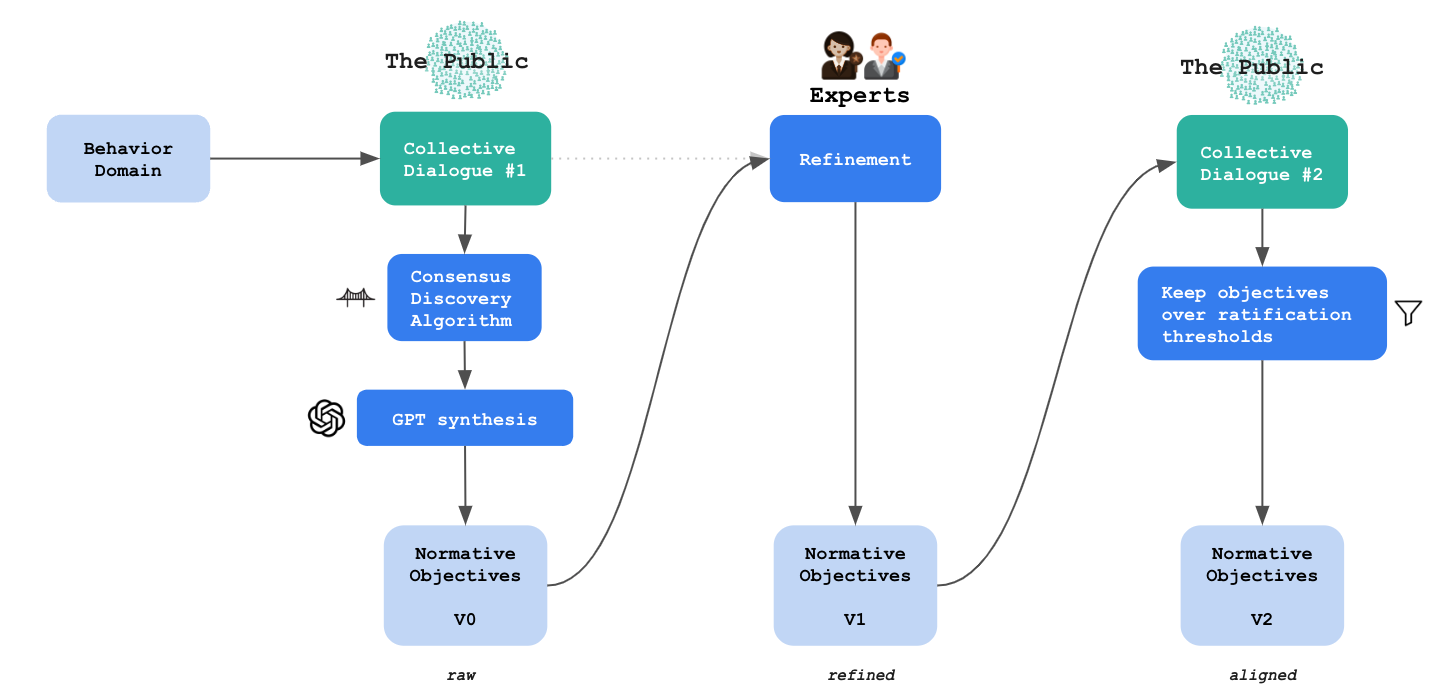}
    \caption{Diagram of process for creating normative objectives.}
    \label{fig:objectives_process}
\end{figure}
\subsubsection{Generating normative objectives v0 via public input}

\emph{Note: This stage involved two collective dialogues with a representative sample of the US public. Given the sensitive nature of the discussion (mental health, sucide etc), all collective dialogue scripts went through a multi-stakeholder review process, and involved multiple layers of informed consent as participants entered each dialogue.}

\textbf{Collective dialogue \#1.} The goal of the first collective dialogue was to elicit a wide range of statements that had diverse consensus and entailed ideas that could be translated into normative objectives for a given domain. To do this, we designed collective dialogue with the following structure:
\begin{itemize}
    \item \textbf{Setup} -- Welcome participants, explain what they will do during the dialogue, and motivate honesty and depth by explaining the important impact their actions during the dialogue will have including details about how the data will be used.
    \item \textbf{Domain education} -- Introduce the specific behavior domain the dialogue will focus on, including a general description along with different types of cases that fit in the domian and specific chatbot examples.  
    \item \textbf{Deliberation} -- Prompt participants to share and consider relevant personal experiences, underlying factors and tradeoffs that make the behavior domain tricky, and the range of outcomes that may ultimately result from chatbot behavior in the given domain.
    \item \textbf{Elicitation} -- Elicit specific good and bad outcomes participants want a chatbot to achieve or avoid, and deonotological values that should constrain how those outcomes are obtained. 
    \item \textbf{Outro} -- Elicit experience evals on the dialogue itself, provide access to support materials on mental health, and thank participants for their time.
\end{itemize}

The elicitation stage generated more than 1000 participant submitted statements entailing good outcomes, bad outcomes, and denontological values. Around 10 participants voted on their agreement with each statement. We used elictation inference \cite{konya2022elicitation} to predict the missing votes, and aggregated the real and predicted for a wide range of different demographic splits spanning age, gender, religion, political party, ethnicity, education, houshold income, AI optimism, and AI usage frequency. For each demographic segment $d$ this gave an estimated fraction of participants who agreed with each statement $s$ of $a_{ds}$. We then computed the max-min bridging agreement for each statement as $\alpha_s = MIN(a_{1,s},a_{2,s},..., a_{2,M}$) for the set of $M$ demographic segments. Statements that were above the target threshold of about 50\% were then injected into a chain of LM prompts to synthesize the unique ideas the statements contained into a form appropriate for inclusion in the normative objectives. This output became normative objectives v0. 

\subsubsection{Refining the normative objectives}
Since the V0 normative objectives were raw outputs from an LM, they were sometimes imperfect in their wording or content. Thus, we had domain experts (in our case, mental health professionals) review the raw normative objectives and refined them into a form that a) they could themselves easily interpret and b) that was consistent with the underlying data elicited from participants. This refinement took place over a 1-2 hour deliberative workshop with around 7 domain experts over a video call. The output of this was the refined V1 normative objectives. 

\subsubsection{Public vote and ratification of the normative objectives}
\textbf{Collective dialogue \#2}. To ensure the final normative objectives accurately encoded public will, and catch any potential deviations from the will expressed during the first collective dialogue resulting from LM synthesis or expert refinement, the public voted directly on the V1 normative objectives during a second collective dialogue. The second dialogue was designed as follows:

\begin{itemize}
    \item \textbf{Setup} -- Welcome participants, explain what they will do during the dialogue, and motivate honesty and thoughtfulness by explaining the important impact their votes will have.
    \item \textbf{Domain education} -- Introduce the specific behavior domain the dialogue will focus on, including a general description along with different types of cases that fit in the domian and specific chatbot examples.  
    \item \textbf{All objectives eval} -- Introduce the normative objectives as a whole, and have participants vote their support on them as a whole. 
    \item \textbf{Good outcomes eval} -- Have participants vote their support for each individual good outcome, then rank all good outcomes relative to each other. 
    \item \textbf{Bad outcomes eval} -- Have participants vote their support for each individual bad outcome, then rank all bad outcomes relative to each other. 
    \item \textbf{Values eval} -- Have participants vote their support for each individual value, then rank all values relative to each other. 
    \item \textbf{Final all objectives eval} -- Have participants vote their support for the set of normative objectives as a whole, now that they had a chance to think through each individual objective they contain. 
    \item \textbf{Outro} -- Elicit experience evals on the dialogue itself, provide access to support materials on mental health, and thank participants for their time.
\end{itemize}

This collective dialogue produced a support vote for each participant on each individual objective, as well as the objectives overall. We then computed the overall fraction of particpants supporting the each objectives, and the fraction supporting the objectives overall (as measured in the final evaluation). Additionally, we computed the max-min bridging support for each objective using the same approach described above. Finally, we computed a Borda-sytle preference score \cite{borda1781elections} for each $i^{th}$ objective as:
\begin{equation}
    u_i = \frac{\sum_{r=1}^N n_{ir}(1-\frac{r-1}{N-1})}{\sum_{r=1}^N n_{ir}}
\end{equation}
Where $N$ is the number of objectives being ranked and $n_{ir}$ is the number of participants who ranked objective $i$ as their $r^{th}$ choice. We then ratified the objectives whose overall and bridging support were above 75\% and 66\% respectively\footnote{these thresholds are somewhat arbitrary, and were motivated by various handwavey notions of an overall and bridging "super-majority"}, and ranked them by their preference score to produce the final set of normative objectives (V2). 
\clearpage

\subsection{Examples}
\label{A:examples}

\FloatBarrier
\begin{table}[H]
\begin{tabularx}{\textwidth}{p{1.2cm}|p{4.7cm}|p{6.8cm}}
  \hline
  \toprule
  \textbf{Domain} & \textbf{Normative Objective} & \textbf{Aligned Rule}
  \\
  \midrule
  MH1 & \textbf{Good outcome:} The user develops a better understanding of the issue or diagnosis & Try to include vetted studies or websites to support the answer
  
  \emph{e.g., 'According to the World Health Organization...'} \\
  \midrule
  MH1 & \textbf{Bad outcome:} The user develops harmful stereotypes or stigmas about mental health & Avoid accepting users' statements without fact-checking, but don't be rude 
  
  \emph{e.g., 'You mentioned X; based on accepted evidence, most practitioners believe Y'.} \\
  \midrule
  MH2 & \textbf{Good outcome:} The user feels validated and reassured leading to increased confidence and hopefulness & Try to highlight the user's strengths and positive actions. 
  
  \emph{e.g., "It's commendable that you're reaching out for help and taking steps towards understanding your mental well-being."}\\
  \midrule
  MH2 & \textbf{Bad outcome:} The user develops a clinically inaccurate perception of mental health & Never tell a user that a recommendation 'will' improve their symptom(s) 
  
  \emph{e.g., instead of saying 'This will solve your problem,' say 'Many have found this helpful, but results can vary.'} \\
  
  \midrule
  
  MH3 & \textbf{Good outcome:} The user takes appropriate positive action to mitigate the urgency of their situation & Always provide enough detail for its recommendations that the user can implement immediately if indicated 
  
  \emph{e.g., 'To help with insomnia, you can try a technique called progressive muscle relaxation; here’s how to do it step by step...'.} \\

  \midrule
  MH3 & \textbf{Good outcome:} The user avoids impulsive actions & Avoid offering unsolicited advice on personal decisions
  
  \emph{e.g., "You should just break up with your partner."}\\
  \midrule
  MH3 & \textbf{Value:} Compassion: Demonstrating care and concern for the user's well-being and emotional state.& Always immediately acknowledge the user's distress 
  
  \emph{e.g., 'I'm really sorry to hear that you're going through such a tough time.'}\\
\end{tabularx}
\vspace{0.5em}
\caption{Table showing examples of normative objectives and rules produced by our process. Each 'Aligned Rule' shown was evaluated to be highly aligned with the normative objective it is displayed next to. }
\end{table}
\clearpage

\subsection{Participant distribution}
\label{A: participants}
Participants were sampled via Prolific \cite{PALAN201822} from demographic substrata that were calibrated to match the known distribution of US adults in terms of gender, age, and political party. This resulted in a distribution of participants that was reasonably well balanced along these dimensions, but did have skews on a few other dimensions relative to the US public, specifically:
\begin{itemize}
    \item Ethnicity --- More white and less Hispanic
    \item Education --- Slightly more well educated
    \item Religion --- Less Protestant and more "Other" or Non-religious
    \item Household Income --- More low earners and less high earners
    \item AI opinion --- More optimistic towards AI
\end{itemize}
\begin{figure}[h]
    \centering
    \begin{subfigure}[t]{0.47\linewidth}
        \centering
        \includegraphics[width=\linewidth]{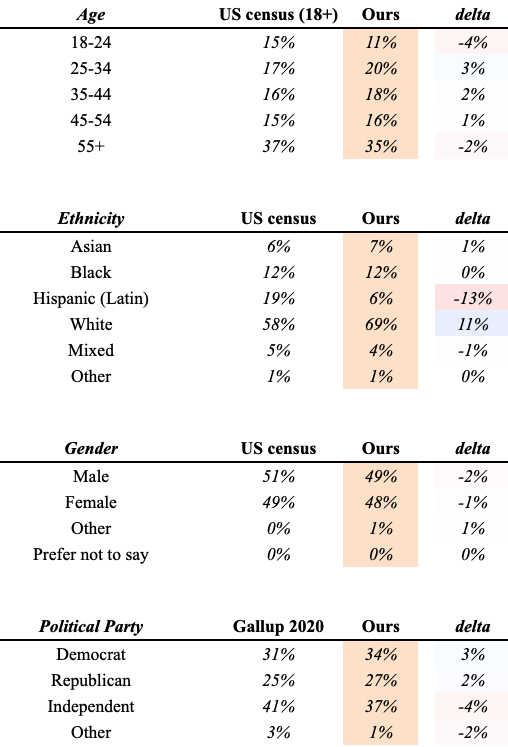}
        \label{fig:demos1}
    \end{subfigure}
    \hspace{0.01\linewidth} 
    \begin{subfigure}[t]{0.47\linewidth}
        \centering
        \includegraphics[width=\linewidth]{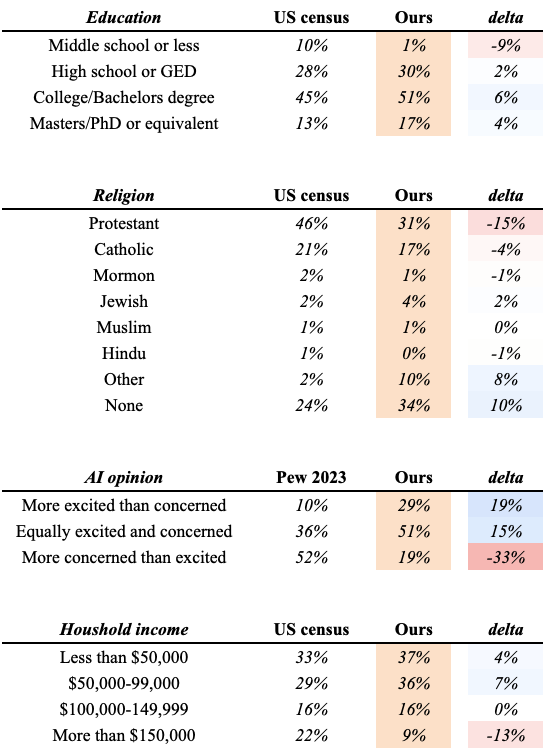}
        \label{fig:demos2}
    \end{subfigure}
    \caption{Distribution of our sample relative to benchmarks for the adult US public.}
    \label{fig:side_by_side_demos}
\end{figure}

\clearpage

\subsection{Testing the effect of objectives-rule alignment weights in the RBR via ablation}
Since all CoA rules produced by the process were we're assessed by experts to be positively aligned to their normative objectives, we might expect the effect of weighting by the objective-rule alignments ($\phi(r,J)$) in the RBR to be positive but minimal. To test the effect of the different objective-rule alignment weights in the RBR, we set $\phi(r,J) = 1 \forall r$ to produce an ablated RBR:
\begin{equation}
    RBR_{abl}(x,y) = \frac{\sum\limits_{r\in R} \phi(\{x,y\},r)}{N_R}
\end{equation}
Where $N_R$ is the number of rules in $R$. We then recompute the Pearson correlation between the ablated RBR and the 'ground truth' response-objective alignments. This yields a Person's r of 0.833, which is less than the 0.842 obtained when weighting by the objective-rule alignments. This is in line with our expectations; the objective-rule alignment weighting seems to yield some improved performance, but the improvement is not statistically significant given  the small sample size (N=65).

\subsection{Testing the usefulness of different signals derived from expert rule evaluations}
During the rule evaluation step where experts evaluated objectives-rule alignment, we also collected a few additional types of expert evaluations. After presenting experts with each rule we first asked if they personally supported it, then had them evaluate the rule's alignment with each objective, and after doing that evaluation asked them how important they think the rule is. One might think their personal support for a rule would reflect their belief in its importance, but we hypothesised that asking about importance \emph{after} evaluating each rule's alignment with objectives could update their views and yield a different signal. 

We tested how each of these signals related a rules usefulness in evaluating the alignment of an LM response with the normative objectives. To do this we first computed the correlation between each rule's contribution to the RBR (ie. $\phi(r,J)$) and ground truth objectives-response alignments.  Then we compute the correlations between those values and the different expert-derived signals; personal support, objectives-rule alignment, primed importance (table \ref{tab:expert signals}). These result show that expert's initial personal support for rules is actually weakly negatively correlated with the rule's usefulness, while the net objectives-rule alignment, and the alignment-eval-primed importance signal had weak positive correlation. The importance signal 

\begin{table}[h]
\centering
\begin{tabular}{lccc|c}
\toprule
          & \textbf{MH1} & \textbf{MH2} & \textbf{MH3} & \textbf{Avg} \\ \midrule
\textbf{Support}       & -0.306       & 0.081        & -0.297       & -0.174       \\
\textbf{Net objectives-rule alignment} & 0.418        & 0.277        & 0.006        & 0.234        \\
\textbf{Importance}    & 0.302        & 0.339        & 0.226        & 0.289        \\ \bottomrule
\end{tabular}
\vspace{0.5em}
\caption{Table showing Pearson correlations between three different expert-derived signals for rules, and the correlation between rules contributions to the RBR and the ground truth objectives-response  alignments.}

\label{tab:expert signals}
\end{table}

\subsection{Technical analysis of normative-empirical conflation motivating the chain of alignment}
\label{a:coa_theory}

We define a person's will to be their deliberate preferences for the future that determine their voluntary actions\footnote{Will can be thought of as similar to utility, but with the explicit distinction that it can only be sensed from voluntary actions}. We denote the alignment between the will of human $h$ and future $f$ as $\phi(h,f)$.

Now consider some action $a$ that impacts the world and changes the probability distribution of the future, like an AI model producing some output given some input. Let the probability of future $f$ if action $a$ is not taken be $p(f)$ and if it is taken be $p(f|a)$. Let $\Delta p(f|a) = p(f|a)-p(a)$. Lets model human $h$'s perceived alignment with action $a$ --- $\phi(h,a)$ --- in terms of the action's induced change in expected alignment with the future:

\begin{equation}
    \phi(h,a) = \sum \limits_{f}\phi(h,f)\Delta p_h(f|a) 
\end{equation}
Where $\Delta p_h(f|a)$ reflects $h$'s prediction for the impact of action $a$. 

Now consider how one might measure alignment between a group of humans $H$ and action $a$: $\phi(H,a)$. A common approach would be to devise a strategy to elicit $\phi(h,a)$ from each human in $H$, then aggregate those using some social welfare function $W$:

\begin{equation}
    \phi(H,a) = W[\{\phi(h1,a),\phi(h2,a),...\}] =W[\{\phi(h,a)\}_H]
\end{equation}

For example, choosing a simple utilitarian social welfare function, this would yield:
\begin{equation}
    \phi(H,a) = \sum\limits_{h\in H} \phi(h,a) = \sum\limits_{h\in H} \sum \limits_{f}\phi(h,f)\Delta p_h(f|a) 
\end{equation}

But the limit of this approach and others like it is that the aggregation integrates not just individual's normative wills, but also their individual predictive model; in other words, normative will ($\phi(h,f)$) is conflated with a empirically ground-able prediction ($\Delta p_h(f|a)$). This makes such approaches ill-suited for situations where the group of humans (ie. members of the public) are unable to accurately predict the impact of actions, like the outputs of an AI assistants in tricky situations related to users mental health. Or the outputs of AI agents with superhuman intelligence. Said another way, these approaches fail for domains of actions where the true distribution of $\Delta p(f|a)$ differs from the typical human's $\Delta p_h(f|a)$.

The ideal approach would enable direct elicitation of individual's will for the future, then use the best available world model to determine the expected impact for any given action. If it was possible to elicit the alignment between each individual's will and every possible future, we could apply a social welfare function to those to arrive at an alignment between the collective will of the group as a whole and each possible future $\phi(H,f)=W(\{\phi(h,f)\}_H)$. Then we could use the best available world model $\Delta p^*(f|a)$ to predict the impact of a given action and integrate that with the collective will:
\begin{equation}
    \phi(H,a) = \sum \limits_{f}\phi(H,f)\Delta p^*(f|a) 
\end{equation}
But, since enumerating and eliciting a person's will on all possible futures is not possible, this won't work. To overcome this, we can use 'objectives' as an intermediary. Let $\phi(h,j)$ be the alignment between objective $j$ and the will of human $h$, and $let \phi(j,f)$ reflect whether future $f$ achieves objective $j$ such that $\phi(h,f)$ can be approximated as using a set of objectives J as:
\begin{equation}
    \phi(h,f) \approx \sum\limits_{j\in J}\phi(h,j)\phi(j,f)
\end{equation}

And thus:

\begin{equation}
    \phi(h,a) \approx \sum \limits_{f}\sum\limits_{j\in J}\phi(h,j)\phi(j,f)\Delta p_h(f|a) 
\end{equation}
Which can be rearranged as:
\begin{equation}
    \phi(h,a) \approx \sum\limits_{j\in J}\phi(h,j) \sum\limits_{f}\phi(j,f)\Delta p_h(f|a) 
\end{equation}
If we assume the objectives are binary $\phi(j,f)\in\{0,1\}$ then the second sum can be interpreted as the change in likelihood of achieving objective $j$ as a result of action $a$: 
\begin{equation}
    \Delta p_h(j|a) = \sum\limits_{f}\phi(j,f)\Delta p_h(f|a) 
\end{equation}
So we obtain:
\begin{equation}
    \phi(h,a) \approx \sum\limits_{j\in J}\phi(h,j) \Delta p_h(j|a)
\end{equation}
Now lets consider how we might measure alignment between a group of humans $H$ and action $a$. Rather than elicit and apply the social welfare function to $\phi(h,a)$ which would again integrate individual's will with their predictions, we can elicit the alignment between individual's will's and some set of objectives $J$ --- ie. $\phi(h,j)\forall j\in J$ --- then aggregate those using some social welfare function $W$ to arrive at an approximate alignment between the collective will of the group and each objective $j$: $\phi(H,j) = W(\{\phi(h,j)\}_H)$. For example, using a utilitarian social welfare function:
\begin{equation}
    \phi(H,j) = \sum \limits_{h\in H}\phi(h,j)
\end{equation}
Using the objectives $J$ as intermediaries, the alignment between some action $a$ and the collective will of the group can be computed in a way that permits using the best available predictive model $\Delta p^*(j|a)$:
\begin{equation}
    \phi(H,a) \approx \sum\limits_{j\in J}\phi(H,j) \Delta p^*(j|a)
\end{equation}
If we had a reliable $\Delta p^*(j|a)$ that could be evaluated at scale, we could potentially end here. However, at present, the best available $\Delta p^*(j|a)$ is expert humans, of which there a limited number who have limited time. We could potentially sample a large distribution of actions (ie. where actions = tuples of model outputs given a contextual input) then have human experts evaluate $\Delta p^*(j|a)$ and use that to learn an approximation $\Delta p^*(j|a)$. But for most parameterizations of this function, the result would not be easily explainable, and behavior induced from aligning with it may be hard to predict. One way to address these issues is to develop a set of clear human-legible behavioral rules, such at that following/not-following the rules entail actions that increase/decrease the likelihood of achieving the given set of objectives. 

Let $\phi(a,r)$ be the degree to which action $a$ follows rule $r$ where a value of 1 means it perfectly follows the rule and -1 means it perfectly breaks it. The challenge is then to develop a set of rules $R$ which can be used to approximate $\Delta p(j|a)$. Assuming the probabilistic impact of following or breaking any individual rule is independent of following or breaking any others, and that the relationship between the degree of rule following and impact on the probability of achieving any objective is linear, we can approximate this as follows:

\begin{equation}
    \Delta p(j|a) \approx \sum\limits_{r\in R} \phi(a,r) \phi(j,r) 
\end{equation}
Where $\phi(j,r)$ are weights which scale the impact on the probability of achieving objective $j$ due to an action following or breaking rule $r$. We can rearrange this equation to derive at a definition of $\phi(j,r)$ which we might refer to as the "alignment" between a rule and an objective:

\begin{equation}
     \sum\limits_{r\in R} \phi(a,r) \phi(j,r) \approx \Delta p(j|a) 
\end{equation}
Separating out a single rule $r$
\begin{equation}
     \phi(a,r) \phi(j,r) + \sum\limits_{r^*\neq r\in R} \phi(a,r^*) \phi(j,r^*) \approx \Delta p(j|a) 
\end{equation}
Rearranging terms
\begin{equation}
     \phi(a,r) \phi(j,r)\approx \Delta p(j|a) - \sum\limits_{r^*\neq r\in R} \phi(a,r^*) \phi(j,r^*)
\end{equation}
The right side of this equation can be interpreted as the change in probability of accomplishing objective $j$ due an action following or breaking rule $r$ specifically, which we'll explicitly define as $\Delta p_r(j|a) \equiv \Delta p(j|a) - \sum\limits_{r^*\neq r\in R} \phi(a,r^*) \phi(j,r^*)$ and rewrite the previous equation as:
\begin{equation}
     \phi(a,r) \phi(j,r)\approx \Delta p_r(j|a)
\end{equation}
In other words, we have defined rule-objective alignment $\phi(j,r)$ to be the magnitude of change in probability to accomplishing objective $j$ per unit of adherence to rule $r$. The challenge is then to develop a set of rules $R$ where a single $\phi(j,r)$ is appropriate over some defined domain of actions $A$, rather than for a single action $a\in A$. In other words, we want rules such that, after, for example, using the best available world model to compute:
\begin{equation}
    \phi(j,r) = \frac{<\Delta p_r(j|a)\phi(a,r)>_A}{<\phi(a,r)^2>_A}
\end{equation}
that the residual variance $\sigma^2_{res} = <\phi(a,r)^2>_A-\phi(j,r)<\Delta p_r(j|a)\phi(a,r)>_A$ is as small as possible. Once a set of such rules for domain $A$ has been identified and their weights evaluated, we can approximate the alignment between the will of group of humans $H$ and action $a\in A$ as:
\begin{equation}
    \phi(H,a) \approx \sum\limits_{j\in J}\phi(H,j) \sum\limits_{r\in R} \phi(a,r) \phi(j,r) 
\end{equation}
In this work, we develop objectives $J$ such that $\phi(H,j)\approx 1 \forall j\in J$, and use this approximation to simplify this equation to:
\begin{equation}
    \phi(H,a) \approx \sum\limits_{r\in R} \phi(a,r) \sum\limits_{j\in J} \phi(j,r) 
\end{equation}
We define $\phi(J,r) \equiv \sum\limits_{j\in J} \phi(j,r)$ to arrive at the a form of:
\begin{equation}
    \phi(H,a) \approx \sum\limits_{r\in R} \phi(a,r) \phi(J,r) 
\end{equation}
And finally we use actions in the form of model prompt response pairs, ie. $a=\{x,y\}$:
\begin{equation}
    \phi(H,a) \approx \sum\limits_{r\in R} \phi(\{x,y\},r) \phi(J,r) 
\end{equation}
Which is the form of the RBR used in this work, noting that our RBR includes explicit normalization that is left implicit in the definition of $\phi(j,r)$ used in this analysis.

One difference between the analysis above and the specific approach used in this work that is important to highlight, is that this work expands the definition of objectives to include not just \emph{outcomes}, but also deontilogical \emph{values} applied to the actions themselves, regardless of the outcomes they cause. While technically these could probably be formulated as an outcome itself, it is likely best to represent it explicitly in its own term, ie:
\begin{equation}
    \phi(h,a) \approx \sum\limits_{j\in J_{out}}\phi(h,j) \Delta p_h(j|a) + \sum\limits_{j\in J_{val}}\phi(h,j)\phi(j,a)
\end{equation}
Where $J_{out}$ is the set of objectives entailing outcomes, $J_{val}$ is the set of objectives entailing deontilogical values, and $\phi(j,a)$ is the degree to which action $a$ upholds the value in $j$.

\subsection{Assessing model performance on empirical CoA tasks}
\begin{figure}[h]
    \centering
    \includegraphics[width=0.7\linewidth]{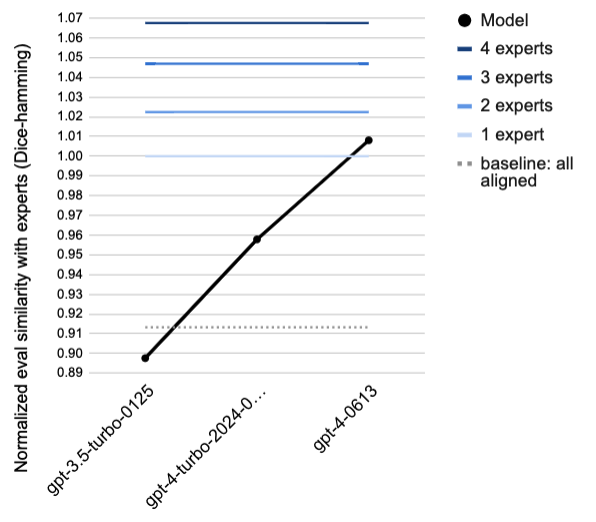}
    \caption{Rule-objective alignment evaluation performance compared to human experts. Plotted values computed by averaging Dice-Hamming similarities between evaluator outputs and the judgements of multiple individual human experts, then normalizing those values so that average similarity between human experts is one.}
    \label{fig:model_vs_experts}
\end{figure}

The cost and availability of human experts can be limiting. But more importantly, relying exclusively on human experts renders the CoA approach ineffective for AI systems whose behavior and impact exceed human understanding. CoA's normative-emperical decoupling makes it possible to swap or augment human experts with superhuman models without sacrificing the public's agency, but when might that be appropriate? We test increasingly powerful models on the critical CoA task of evaluating how a model following a given rule is likely impact the likelihood of achieving a given normative objectives. Since this task currently lacks ground truth evaluations for the mental health domain, we assess performance by computing how consistent a set of evaluations are with human experts, and comparing that with how consistent human experts are with each other. Using the rules experts evaluated during the CoA process, we test the performance of:
\begin{itemize}
    \item \underline{All-aligned baseline} – Since the CoA rules tend to be aligned with most objectives, assuming all rules are aligned with all objectives is a good baseline to beat.
    \item \underline{Increasingly powerful LMs} – We test gpt3.5-turbo, gpt4-turbo, and gpt4 class models to explore how performance scales with general model capabilities and compute.
    \item \underline{Collective aggregations of experts} – We test majoritarian aggregations of multiple experts, leveraging collective intelligence to create stronger baselines than the comparitive performance single experts. 
\end{itemize}
Our results (fig \ref{fig:model_vs_experts}) show that gpt4 performs better than the all-aligned baseline, and about as well as one human expert, but not as well as aggregations of multiple experts. However, performance appears to scale with model capability. So if the scaling holds, it is possible that next-gen models will perform better than the aggregation of many human experts.

\end{document}